# CROSSOVER TO NEGATIVE DIELECTRIC CONSTANT IN PEROVSKITE $PrMnO_3$

SANTHOSH KUMAR BALU[1], N. PRAVEEN SHANKER[1], M. MANIKANDAN[2], N. APARNADEVI[1,4], T. MUKILRAJ[1], P. MANIMUTHU[3], AND C. VENKATESWARAN[1a]

[1] *Department of Nuclear Physics, University of Madras, Guindy campus, Chennai – 600 025, India*
[2]*Centre for Nanoscience and Technology, Anna University, Chennai – 600 025, India*

[3] *Nanostructures and Functional Materials, Centre for Materials Science and Nanotechnology, University of Oslo, Norway*

[4]*Department of Physics, Ethiraj College for Women, Egmore, Chennai, India – 600 008*

[a]*Corresponding author Email: cvunom@hotmail.com*

**Abstract**

Negative dielectric constant materials have the potential for unique optical and microwave applications which will also reduce the complexity in device architecture. We report the observation of temperature dependent Negative Dielectric Constant (NDC) above 503 K in the perovskite $PrMnO_3$, confirmed through impedance spectroscopy, obeying classical Drude theory. The principle behind the associated negative dielectric loss is explained using the Axelrod mechanism.

**Introduction**

In the last three decades, $ABO_3$ perovskites ( A -Tb, Dy, Eu, Pr, Nd, Gd, Bi , B- Mn, Ti, Co, Fe, Mg)  draw huge attention because of their potential in applications like magnetic storage, magnetic switching, fuel cells, electronic switching and in solar cells. Among them, $AMnO_3$ system is known for its structural stability, interesting thermal, electrical and magnetic properties. They exhibit unusual behaviour such as, the field induced metamagnetic transition in $DyMnO_3$ due to spin reversal of Dy-magnetic moment [1], magnetic transition in $GdMnO_3$ due to the spin wave ordering of $Mn^{3+}$ and $Gd^{3+}$[2]. $TbMnO_3$ is a versatile compound in this family associated with spin polarization, spin reversal of $Mn^{3+}$ and $Tb^{3+}$ ions, spin frustration and ferroic ordering – control of magnetic with electrical polarization [3]. $NdMnO_3$ exhibits exchange bias effect due to the correlation in spins of $Nd^{3+}$ and $Mn^{3+}$ with the applied field [4]. Negative dielectric property further compliments the $AMnO_3$ family making them potential for application in antennas.

Generally the force between a system of charges is well described by Coulomb's law as their product and inversely proportional to square of distance the between them, given by $F= Kq_1q_2/\varepsilon_r r^2$ [5, 6] where $\varepsilon_r$ is relative permittivity of the medium. The sign indicates the nature of force: positive for repulsive and negative for attractive, if the force between like charges is to be attractive, the only possible way is $\varepsilon_r$ should take negative value, which physically signifies the alignment of charges opposite to the applied voltage or electric field. This opens a new gate towards NDC which is also termed as negative index materials (NIM) or metamaterials. NIM are man-made matrix material of metallic unit cell which exhibits NDC due to plasma-like effect. The dielectric constant of the material is related to polarization as *$p = \varepsilon_0 (\varepsilon'-1)$* [6-8] where $\varepsilon'$ is the real part of dielectric constant, $\varepsilon = \varepsilon'-i\varepsilon''$ and $\varepsilon''$ is the imaginary part of dielectric constant [9]. NDC represents polarization which opposes the applied field, similar to magnetic field lines expelled by diamagnetic materials [9]. The above condition happens when $k \neq 0$ and $\omega \to 0$ [6, 10].

In the year, 1981 Dolgov et.al proved that NDC is in accordance with the principle of electrodynamics and these stable materials have the possibility to exist [11]. At present, only few reports are available which shows NDC [12-18]. However the underlying physics is not fully understood. Material which displays NDC is also associated with negative dielectric loss (tan δ). In short, when the like charges form a pair in the applied oscillating electric field (NDC), it is capable of giving distinct properties like superconductivity at room temperature, reverse Doppler Effect, reversed Cherenkov radiation [6, 19, 20]. Finding a system with a NDC is a challenging task and very few reports are available. Chu *et.al* found negative dielectric constant in '$Al_2O_3$ in silicon oil system' [21] at very low frequencies., Yan *et.al* found NDC in non-conducting polymers as a function of time [15]. Even negative permittivity has a great potential in wave absorbing, wave transmission in antennas, electromagnetic radiations shields. Further, materials showing NDC as a function of frequency, time and even with respect to the applied field are reported, but as a function of temperature has not been reported yet in $AMnO_3$ systems. Among the $AMnO_3$ compounds, $PrMnO_3$ is less insulating, a property which will increase the number of charge carriers with temperature.

This letter reports the first observation of negative dielectric constant phenomenon in pervoskite a $AMnO_3$ system as a function of temperature. The observed NDC is also associated with a change in resistive-capacitive to resistive-inductive nature.

**Experimental**

$PrMnO_3$ is prepared by initially milling the stoichiometric ratio of $Pr_6O_{11}$ and $MnO_2$ in zirconia vial with a ball to powder ratio of 27:11 for 5 h (300 rpm) and the black powder obtained is pressed into a pellet of 12 mm diameter under a pressure of 5 tonne and heat treated at 1473 K for 5 h in a box furnace. The SAED pattern and surface morphology are taken in a high resolution transmission electron microscope (TECNAI-G2, FEI-Neitherland) and Field emission scanning electron microscope (FE-SEM Hitachi SU6600), respectively. $PrMnO_3$ pellet of 8 mm diameter and 1.25 mm thickness, is subjected to AC impedance spectroscopy study in a Solatron 1260 impedance analyzer (with Pt electrode), a key tool in understanding the electrical property of materials. In impedance spectroscopy, the resulting current is a function of sinusoidal wave, $V=V_0 \sin(\omega t)$, from which the Fourier transformation of the applied wave is resolved as $Z(\omega)=V(\omega)/I(\omega)$, where Z is impedance of the sample. The impedance part $Z(\omega)$ can be resolved as $Z(\omega) = Z'(\omega) + Z''(\omega)$ [9]. Z" (imaginary part) versus Z' (real part) is collectively called the Nyquist plot which confirms capacitive, inductive and resistive nature of the material. First quadrant of Z' versus Z" (upper semicircle) confirms resistive-inductive behaviour and the fourth quadrant confirms resistive-capacitive behaviour. An R-L-C circuit produces a full circle distributed in both these first and fourth quadrants [8, 13].

**Results and discussion**

Phase purity of the prepared sample is confirmed by indexing the selective area electron diffraction (SAED) pattern (Fig. 1.a) of the sample, and the FE-SEM micrograph confirms a uniform growth of the grains which is a result of high energy ball milling followed sintering at high temperature. Fig. 2 (a) shows schematic diagram of impedance analysis and Fig. 2 (b, c and d) shows the plot of real and imaginary parts of impedance as a function of temperature from 303 K to 503 K. The Nyquist Plot (Z' versus Z") showing an intersection on the *x*-axis (fitting it to a circle) provides the resistance value of the prepared $PrMnO_3$ sample. Fig. 2 (b) shows the resistance at room temperature (303 K) in the order of kΩ. As the temperature is increased, the resistance of $PrMnO_3$ decreases, which is confirmed by the decrease in the radius (Fig 2.c) of the semicircles [9]. Therefore, sample shows a wide range of resistance from 30 kΩ at 303 K and 450 Ω at 483 K. The occupancy of Z" and Z' in the fourth quadrant confirms the presence of resistive-capacitive nature of the sample [8, 13].

As the temperature of the sample increases, the resistance decreases abruptly, as shown in Fig. 2 (d) [3]. The transformation of the semicircle from fourth quadrant to first quadrant (Z' vs Z") confirms the presence of inductive nature in the sample (resistive-inductive) with respect to the increase in temperature [13,15]. The graph is attributed to the competition between capacitive and inductive behaviour. The temperature at which the behavior completely switches from resistive-capacitive to resistive-inductive can be defined as the capacitive-inductive temperature ($T_{CI}$ ~ 503 K). If T < $T_{CI}$ – sample exhibits resistive-capacitive nature and if T > $T_{CI}$, it behaves as resistive-inductor. Thus PrMnO$_3$ exhibits resistance-capacitive nature below $T_{CI}$ and resistive-inductive nature above $T_{CI}$.

The conductivity plot as a function of temperature (Fig 3.a) which confirms the semiconducting behaviour throughout the temperature regime. The σ (T) plot is fitted with normal exponent (MB statics) i.e., $\sigma = \sigma_0 \exp(-ib/T)$, where σ$_0$ and b are constants. The fit shows a good agreement with obtained data and the insert shows the residual value of fitting with $R^2$ value fitting value of 0.983. Thus, from the conductivity graph, we conclude that PrMnO$_3$ obeys classical theory. Fig. 3 (b) shows the dielectric plot as a function of frequency and temperature. It is noted that at room temperature, the dielectric plot is normal in behaviour with the frequency. As the temperature is increased, the dielectric constant (ε$_r$) decreases, but as the temperature is increased above $T_{CI}$, the ε$_r$ becomes negative value at certain temperature and frequencies. This is significant according to classical Drude theory [6, 22]. The negative value in dielectric constant can also be attributed to, electric field in the sample is opposite to that of the applied electric field (i.e.) the polarization inside the sample is in the direction opposite to the external /applied electric field. The NDC can also be termed as dia-electric property of sample [21]. The observed phenomenon is also similar with repelling behaviour of diamagnetic material in external magnetic field and formation of cooper pair electrons in superconductor below $T_c$.

Using Fourier transformation (AC–impedance spectroscopy), dielectric component of a dielectric material can be resolved as $\varepsilon_r = \varepsilon' - j\varepsilon''$, where $\varepsilon_r$ is resolved as real and complex parts, respectively [9]. According to Drude theory, the negative dielectric constant can only happen if electron-electron combines by overcoming their repulsion (*i.e. Columbic force should be attractive*) or forming an electron gas [6, 10]. Chu.et.al also describe negative dielectric constant with the term dia-electricity [21] which can happen only in the low frequency regime due to the wave-vector as discussed in introduction section.

Inset of Fig. 3(b) shows the resistivity plot. Figure 3(c) shows imaginary part of dielectric constant as a function of frequency and temperature. Imaginary part ($\varepsilon''$) is inversely proportional to frequency. At high frequencies the magnitude is low and vice versa. The imaginary part of the dielectric constant is also related to conductivity of the material as, $\sigma_d = \omega \varepsilon_0 \varepsilon''$ [6, 8]. The electric modulus M' vs M" is plotted (Fig. 3.d) which is more significant to temperature transitions [9] shows a clear transition of quadrant from 1$^{st}$ to 2$^{nd}$, which is due to capacitive – inductive transition. In this, a change with respect to temperature is an evidence for negative dielectric constant in PrMnO$_3$ system.

Fig. 4 shows tan δ plot at 1 kHz and 10 kHz. The positive value of tan δ (dielectric loss) is due to normal relaxation. But the negative value is unusual. In 2006, Axelrod et al. proved that negative tan δ is due to the release of more energy than absorbed by the sample [23]. However according to the principle of conservation of energy, the total energy is always a constant and there must be a mechanism to store the energy and to release out at a critical temperature and frequency [23]. DC conductivity (log σ vs 1000/T) plot shows two thermally activated regions and the activation energy is obtained by fitting the data using the Arrhenius equation $\log \sigma = \log \sigma_0 - \frac{E_a}{kT}$, where $\sigma_0$ is the pre-exponential factor, E$_a$ is the activation energy and $k$ is Boltzmann's constant. The activation energies from the fitted data are 0.18 eV and 0.09eV with R$^2$ values of 0.977 and 0.979 respectively.

**Conclusion**

To conclude, pure PrMnO$_3$ prepared by ball milling assisted heat treatment reveals resistive - capacitive to resistive - inductive transition (T$_{CI}$) at 503 K. Negative dielectric constant in the low frequency regime beyond T$_{CI}$, as a function of temperature, is reported in PrMnO$_3$, a property which provokes interest to explore the material further for micro-optical device applications.

**Acknowledgment**

BSK (Santhosh Kumar Balu) thank DST (Department of Science and Technology) – Inspire (Innovation in Science Pursuit for Inspired Research) for the award of Senior Research Fellowship (SRF-IF-140582) and Mr. B. Soundararajan for his support in the measurements. NCNSNT, University of Madras, is gratefully acknowledged for FE-SEM, SAED measurement.

**Conflict of interest**

The authors declare that they have no conflicts of interest to this work.

**Figures and Captions**

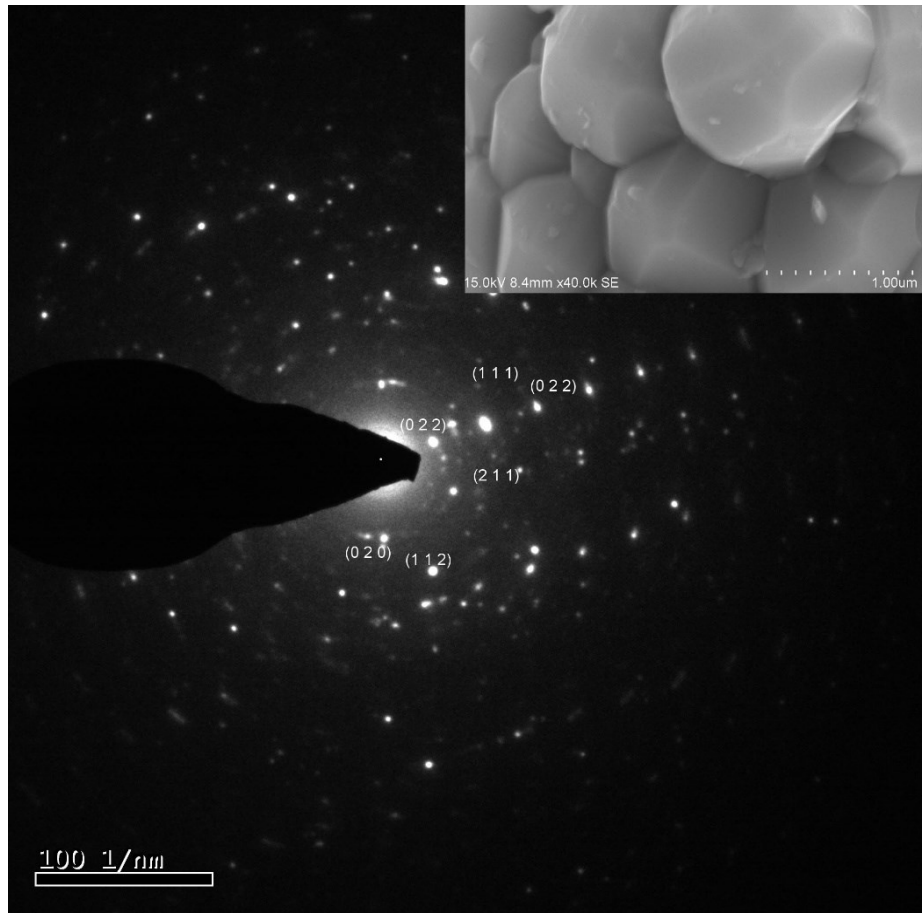

**Fig. 1.** Selective area electron diffraction (SAED) showing the characteristics reflections (0 2 2), (0 2 0), (1 1 2) and (0 2 2) of orthorhombic $PrMnO_3$. FE-SEM (insert) evidences a uniform grain distribution that is due to ball milling and high temperature heat treatment.

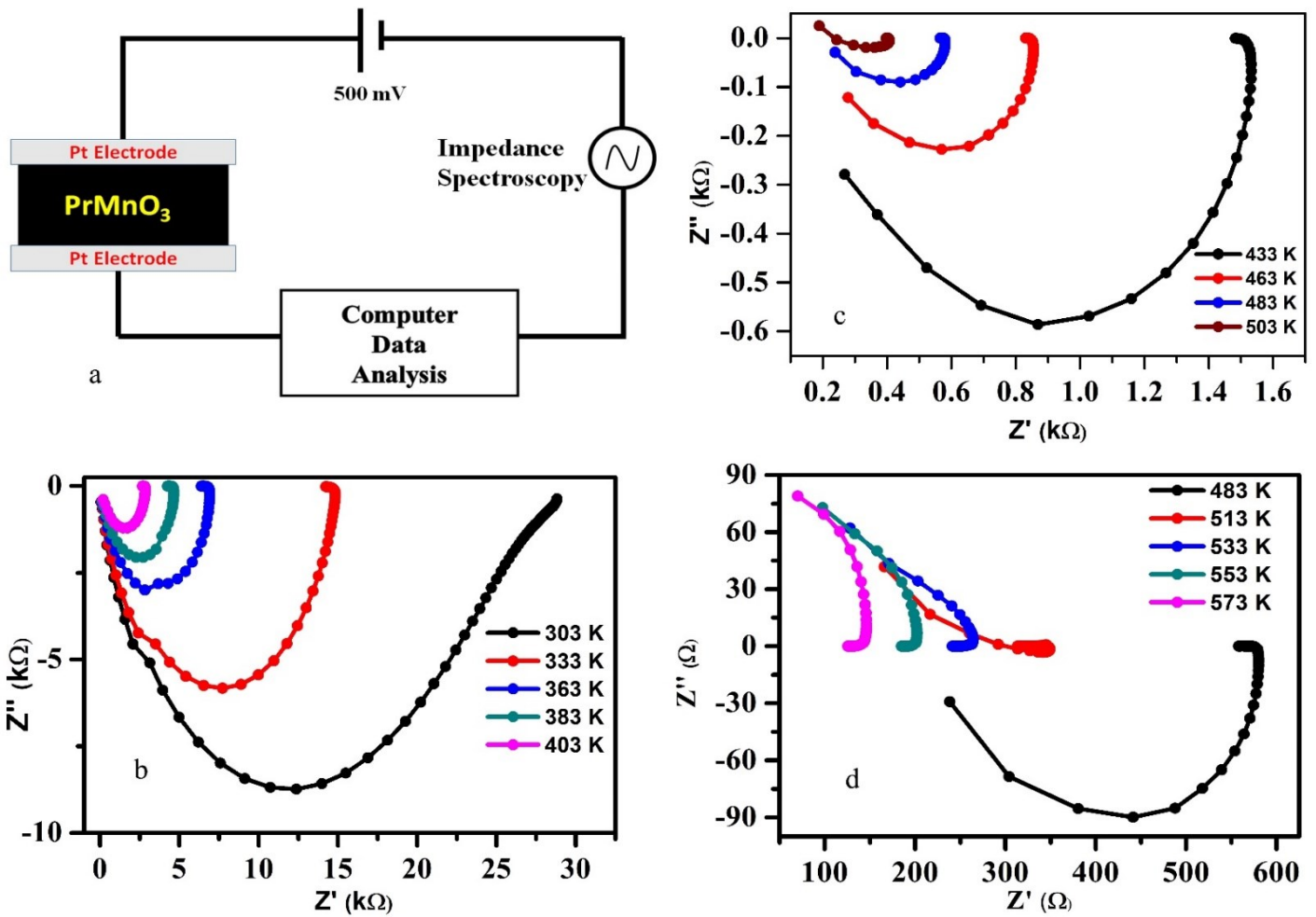

**Fig. 2.** (a) Shows schematic of impedance measurement carried out by placing PrMnO$_3$ between two platinum electrodes maintained at a constant applied DC potential of 500 mV. (b) Real (Z') versus imaginary (Z") plot of impedance in the temperature range of 303 K to 403 K. The trend of decreasing semicircles confirm the metallic behaviour. (c) One fold decrease in the magnitude of Z' Vs Z" between 433 K and 503 K (compared to the magnitude in the range of 303 K to 403 K) further confirms the metallic nature (d) The shift of semicircles from the 4th quadrant to 1st quadrant confirms the inductive nature of PrMnO$_3$ and the order of magnitude is three fold at low temperature.

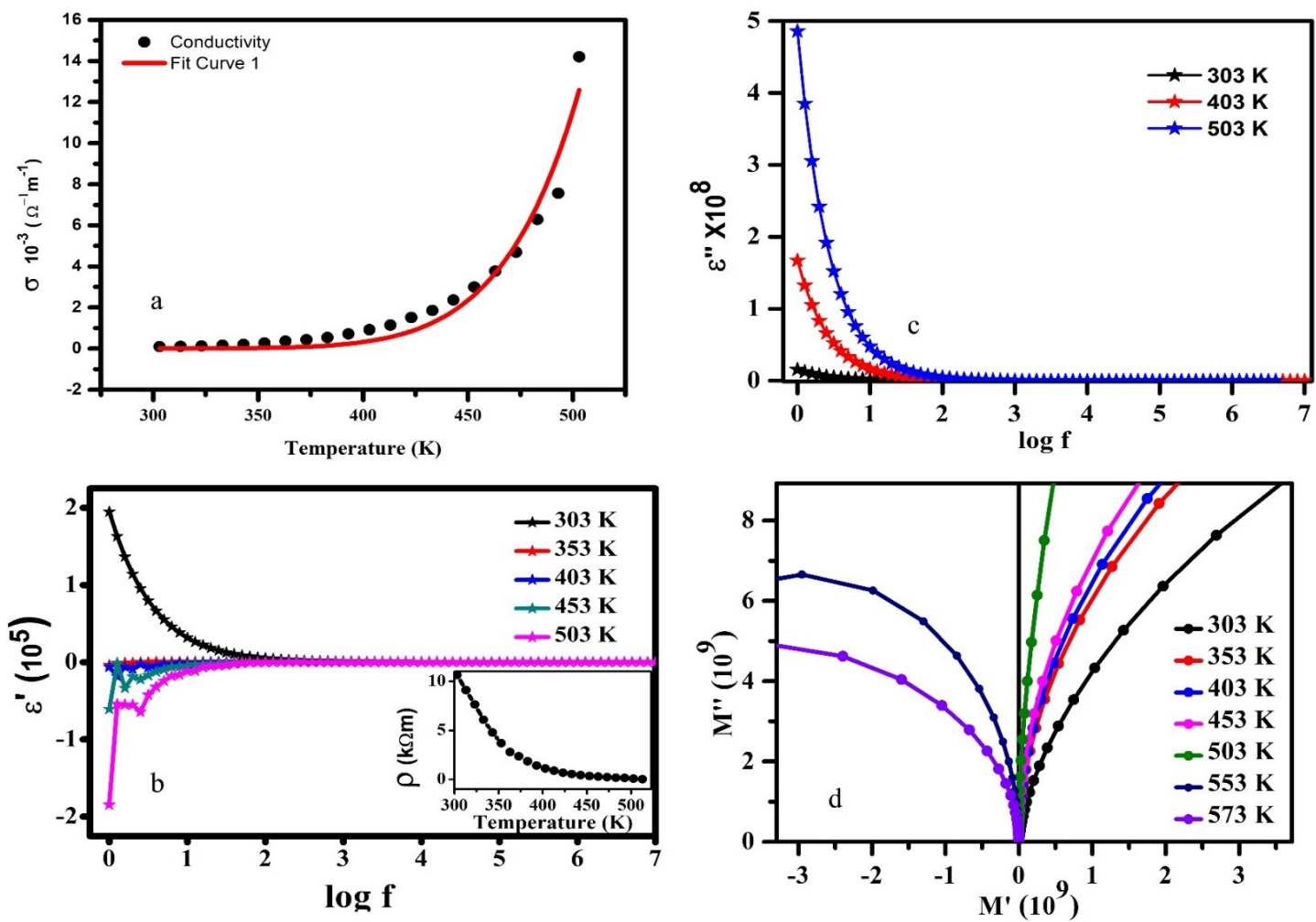

**Fig. 3.** (a) Conductivity versus temperature plot fitted using the Maxwell's Boltzmann (MB) equation $\sigma(T) = \sigma_0 e^{(-ib/T)}$. Inset shows residue of the MB fit $R^2$ value of fitting 0.983 (b) Dielectric constant as a function of temperature in the range 303 K – 503 K has a magnitude of $10^5$. Inset shows the resistivity of PrMnO$_3$. (c) Increasing trend is observed in the Loss spectrum of PrMnO$_3$ at various temperatures and the graph shows positive value throughout the temperature and frequency (d) Modulus spectra (M' vs M") show a clear evidence of change in quadrant from 2$^{nd}$ to 1$^{st}$ which is due to inductive property.

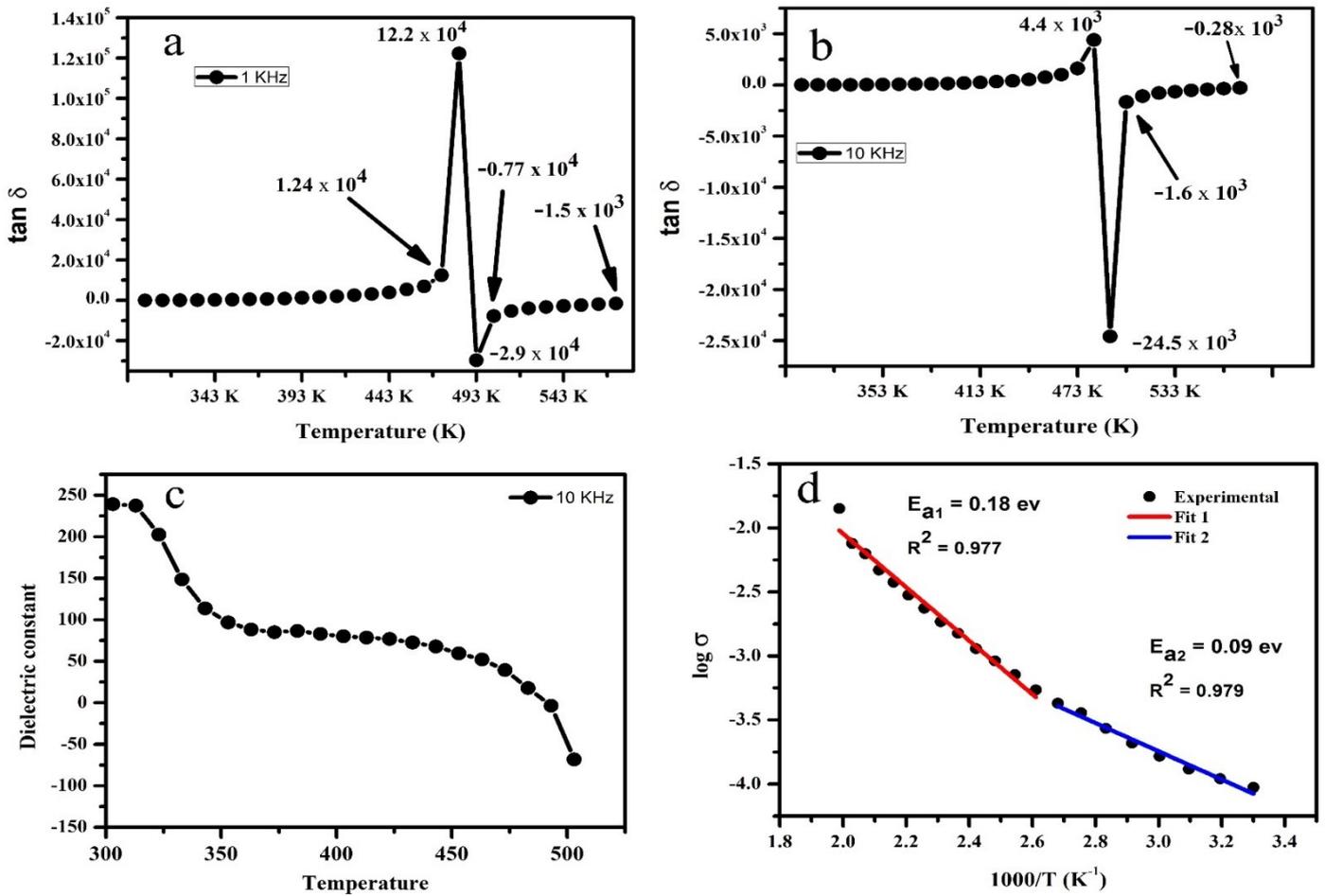

**Fig. 4.** (a) and 4(b) shows tan δ (dielectric loss) of the sample at 1 kHz and 10 kHz. The value of tan δ reaches a maximum value and then shifts to negative scale beyond 503 K remain in negative. The dielectric constant of $PrMnO_3$ at 10 kHz (fig 4.c), the sample shows negative value beyond 503 K. Arrhenius plot showing (log σ vs 1000/T) two thermally activated region of the sample. The data is fitted with straight line form of Arrhenius and the corresponding activation energy is shown.

# Supplementary

# Instrumentation

Impedance analysis of the sample was carried out in a Solatron 1260 impedance analyser which is known for its instrument resolution and accuracy.

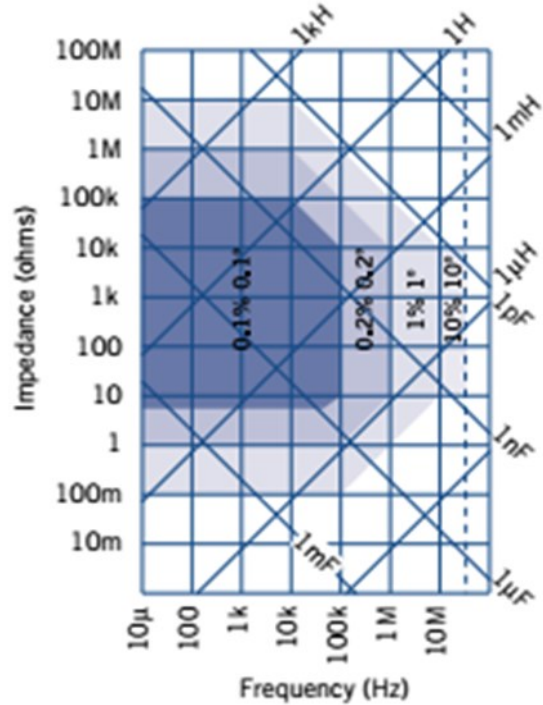

**Fig. 1**. Impedance plot as a function of frequency.

Source: https://www.upc.edu/sct/en/documents_equipament/d_112_id-537.pdf).

The plot confirms, even at a frequency of 100 Hz, the instrument measures 1 Ω, with a least capacitance of 1 nF with an accuracy of 0.2% .

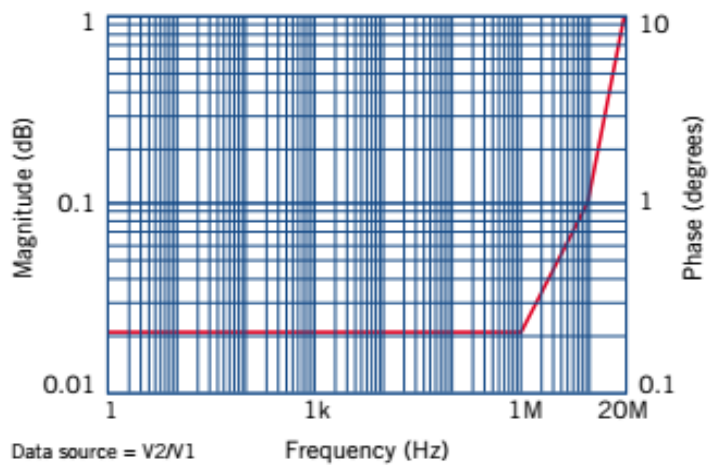

**Fig. 2.** Frequency as a function of phase angle. (Source: https://www.upc.edu/sct/en/documents_equipament/d_112_id-537.pdf)

## Structural analysis

The prepared sample was subjected to X-ray diffraction (XRD) analysis for phase confirmation. The lattice parameters calculated agree with JCPDS file no. 072-0377.

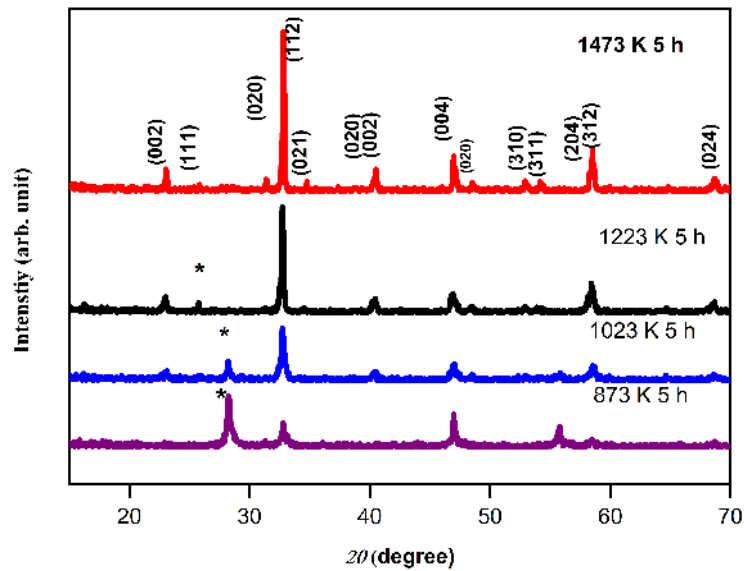

**Fig. 3**. XRD pattern of the prepared PrMnO$_3$ sample.

## Reinforcement of inductive nature of the sample

1) In principle, if a dielectric medium of a capacitor is associated with negative dielectric constant (NDC), then the capacitor is equivalent to inductor due to which inductive nature is also present. (i.e.) a capacitor filled with negative dielectric constant is equivalent to inductor.

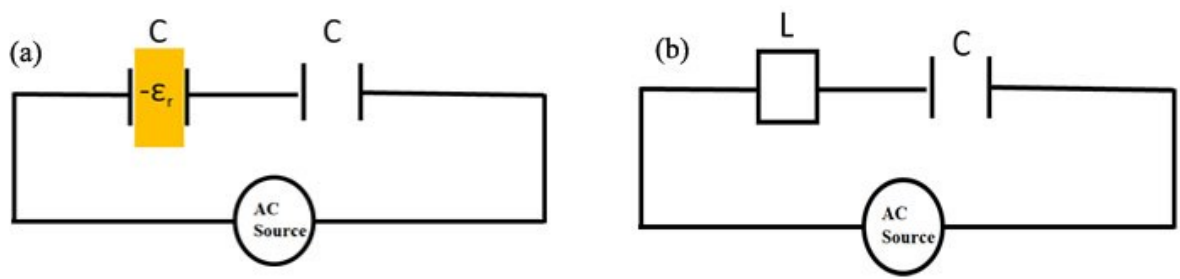

**Fig. 4**. *Schematic diagram showing (a) CC' circuit showing C' is filled with material which exhibit NDC effect (b) LC circuit with a.c. source connected. In priciple with negative index materials both the circuit are same.*

The prepared PrMnO$_3$ shows negative dielectric constant as well as a transformation of semicircle in Z' vs Z'' plot, which strongly confirms the inductive nature of the sample. Therefore, the observed NDC is also in line with the priciple of negative index materials.

(Ref: Physics of negative index materials – S. Anantha Ramakrishna, Rep.Progs. Phys 68, 449-521, 2005. 2. Book on Physics and applications of negative refractive index materials- S. Anantha Ramakrishna, Tomasz M. Grzegorczyk, SPIE press- CRC press.).

2) In order to prove that the observed inductive effect and NDC is due to sample and not because of platinum probe, the data obtained at 573 K is fitted using the equivalent model (CR(LR)) which gives the inductive value of 38.4(3) mH with $\chi^2$ = 1.2E-3. Also L is parallel with C, which confirms that NDC is because of sample.

3) Using the expression (Ref: Handbook of chemistry and Physics, 44[th] Ed, Chemical Rubber Publishing Co, Cleveland, OH, 1962.)

$$L = 2l \left[ 2.303 \log \left\{ \frac{4l}{d} - 1 \right\} + \frac{\mu}{4} + \frac{d}{21} \right]$$

Where

    Diameter (d)        = 0.1 mm
    Wire length (l)     = 5 cm
    Permeability (µ)    = 1.256E-6 H/m

By substituting the values, the inductance is calculated as 66.01(3) nH. (Self-inductance of Pt probe). Thus L value calculated (because of Pt probe) is nearly six times lesser than L of sample (i.e.) $L_{wire}$ <<< $L_{sample}$.

4) Also, We would like to point out the effect of stray inductance in the impedance studies and how we have overcome it.
   - The stray inductance component of Pt wire will have considerable effect in the plot of Z' and Z" if the magnitude is less than 2Ω. Otherwise the effect is negligible.
   (Ref: S.Primdahl and M.Mogenson-J.Electrochem. Soc. Vol 144, Pg 3409-3419, 1997).
   - The contibution due to working electrode, counter electrode or refrence electrode will have a Z' and Z" spectra if only the freqency is greater than 10 MHz. At this point, the subraction of stray inductace component will have reasonable consideration in the interpretation of data at the higher frequencies (~10 MHz).
   (Ref: 1. Nickel/YTTRIA-stabilized zirconia cermet anodes for solid oxide fuel cells- Soren Primdahl – 1967 – Thesis.

2. Oxidation of hydrogen on Ni/YTTRIA- Stabilized Zirconia cermet anodes – S.Primdahl and M.Mogensen - J.Electrochem.Soc – Vol 144,1997.)

➢ We would like to putforth one more example from the book of Impedance spectrosopy theory, experiment and applications, 2nd Ed, edited by Evengenij Barsoukov and J.Ross Macdonal (wiley –Interscience) From chapter 4 :

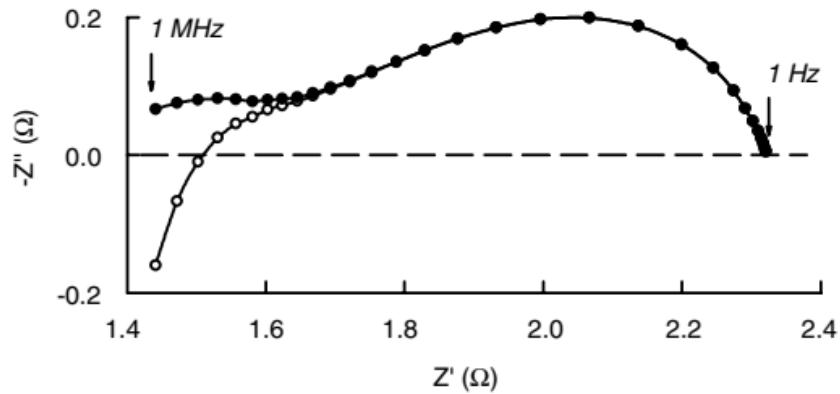

**Figure 4.1.24.** Effect of inductance error on the impedance spectrum of a symmetrical cell. Open symbols: raw data; closed symbols: data corrected for stray inductance of $36 \times 10^{-9}$ H. The sample is a YSZ tape of area $0.21\,cm^2$ with an LSM composite cathode on both sides, measured in air, 850°C.

The figure confirms, the stray inductance will have effect (1) at higher frequencies and (2) the magnitude of Z" should be less than unity.

5) At last, we have fitted the data at 573 K, by including the stray components. The data is fitted with two models (a) $L_2(R_1L_1)$ (b) $R_1L_1$

   Where $L_2$ – inductance due to stray components ($6.6 \times 10^{-8}$ H)

   $R_1$ – Resistance (146 Ω)

   $L_1$ – inductance due to sample (39 mH)

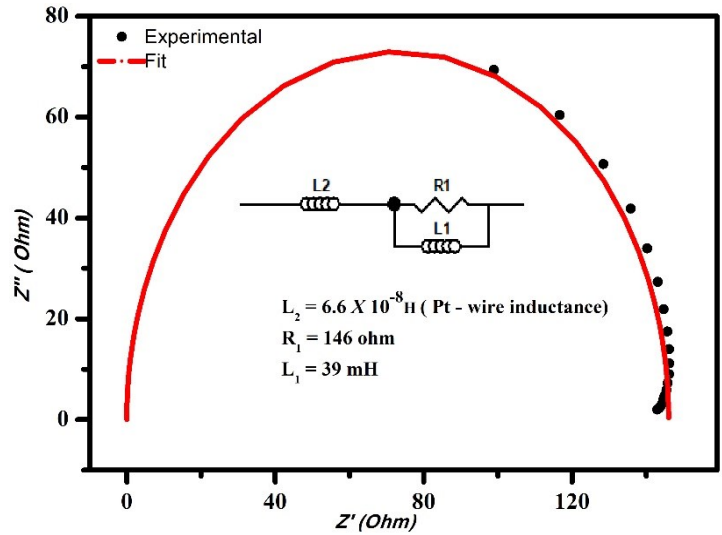

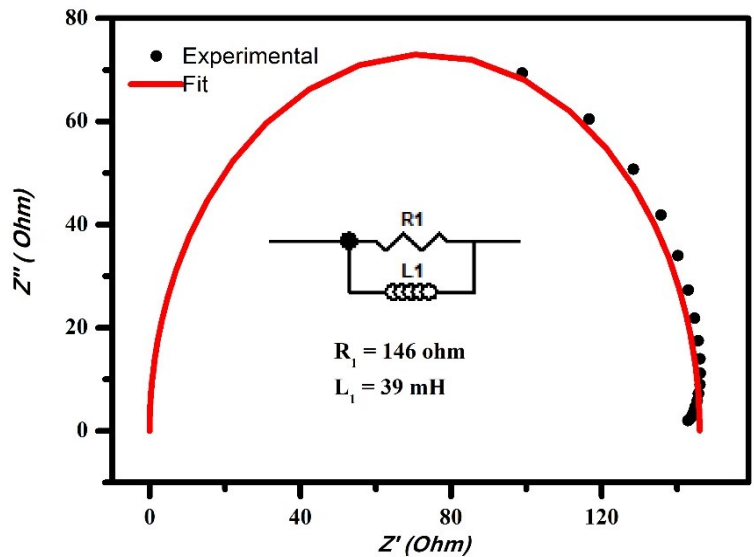

The above diagram shows the stray component variation with the measurements.

The table below, for same Z' the variation in Z", with and without stray components.

| Frequency (Hz) | Z' (Ω) | Z"(Ω) With stray components | Z"(Ω) Without stray components |
|---|---|---|---|
| $10^6$ | 145.999 | 0.50168 | 0.08699 |
| $10^5$ | 145.995 | 0.89249 | 0.85005 |
| $10^4$ | 145.528 | 8.2848 | 8.28052 |
| $10^3$ | 111.519 | 62.010 | 62.0098 |
| $10^2$ | 4.7826 | 25.988 | 25.9884 |
| $10^1$ | 0.0342 | 2.23431 | 2.23431 |

This confirms, only when $\upsilon \to \infty$ (higher frequency ~ $10^6$ Hz) the stray components have considerable effect. From the basic principles of physics, the inductance varies with frequency by $X_L = L(2\pi\upsilon)$. The magnitude of NDC observed in PrMnO3 sample is considerably high in the lower frequency regime when compared to that in the higher frequency regime. The figure below shows the response of Pt probe (shorted) for a frequency sweep of 1 Hz to $10^7$ Hz. It can be clearly noticed, at lower frequencies (less than $10^5$ Hz) the plot of Z" vs Z' shows a flat response and at $10^7$ Hz the Pt probe(s) offers a resistance of ~2.4 Ω which is very less than the sample resistance.

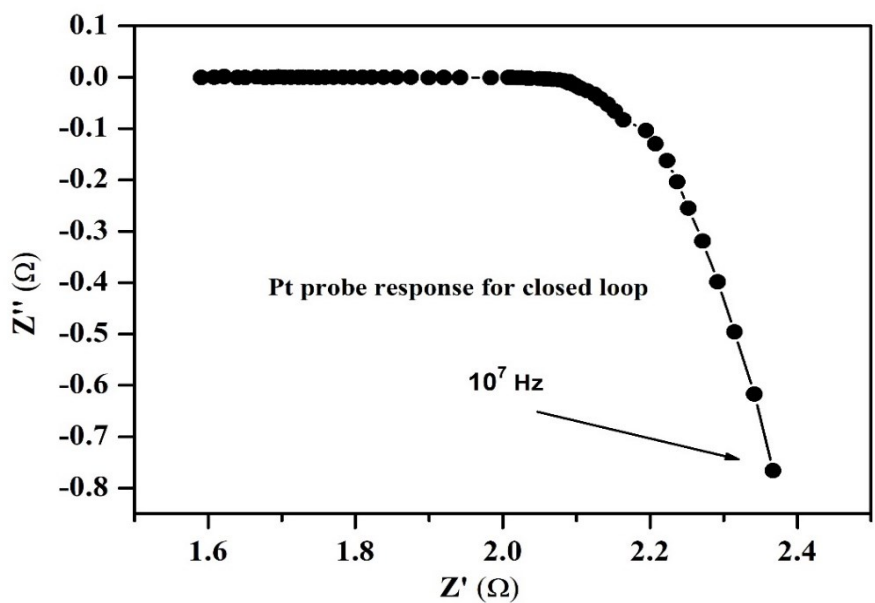

6) The four probe (van der pauw method) experiment (to compare the resistivity value obtained from impedance spectroscopy) was performed using Keithley source meter and the potential difference was measured using Keithley nano-voltmeter.

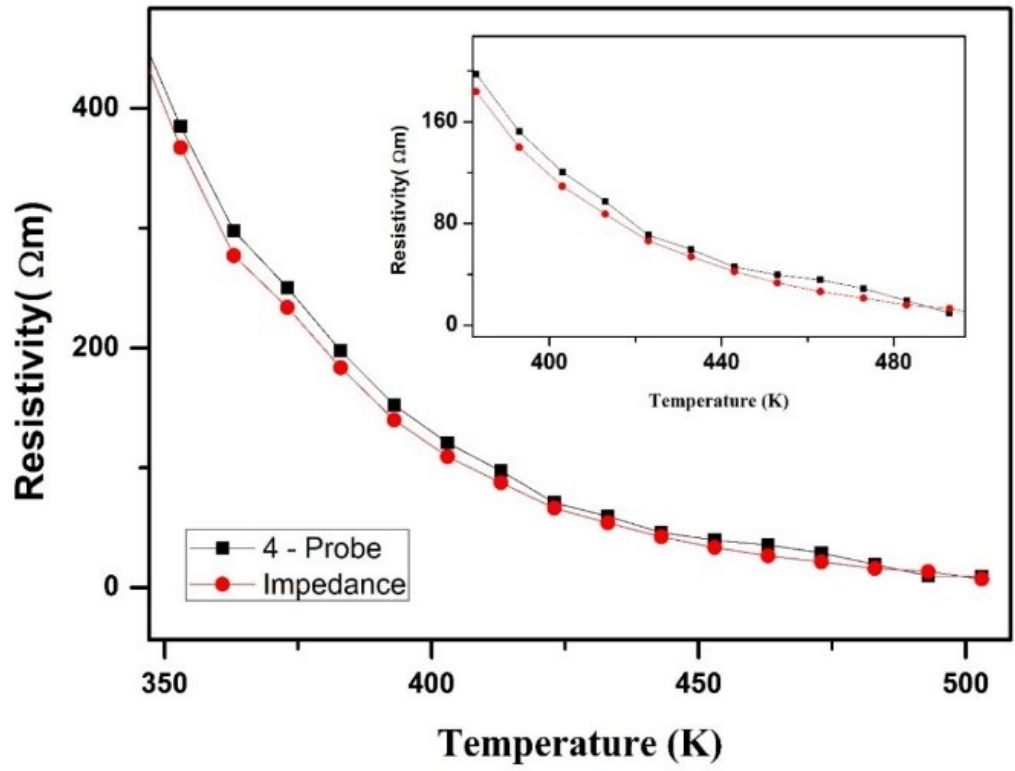